\newcommand{\eqn}[1]{\label{#1}}
\newcommand{\eq}[1]{\begin{equation}#1\end{equation}}
\newcommand{\eqs}[1]{\begin{eqnarray}#1\end{eqnarray}}
\newcommand{\br}[1]{\left(#1\right)}
\newcommand{\bra}[2]{{}_{#2}\big{<}#1\big{|}}
\newcommand{\ket}[2]{\big{|}#1\big{>}_{#2}}
\def\lb{\nonumber\\}%linebreaking without eqn #
\newcommand{\refbr}[1]{(\ref{#1})}

\documentstyle[12pt]{article}

\def\a{\alpha}

\def\g{\gamma}
\def\d{\delta}

\def\l{\lambda}
\def\m{\mu}

\def\G{\Gamma}

\def\/{\over}
\def\*{\partial}
\def\|{\mid}

\def\hatpsi{\hat{\psi}}
\def\barS{\bar{S}}
\def\barU{\bar{U}}
\def\barM{\bar{M}}

\def\barpsi{\bar{\psi}}
\def\hatW{\hat{W}}

\newcommand{\NP}[1]{ Nucl. Phys.\ {\bf #1}\ }
\newcommand{\PL}[1]{Phys. Lett.\ {\bf #1}\ }
\newcommand{\CMP}[1]{Comm. Math. Phys.\ {\bf #1}\ }
\newcommand{\LNC}[1]{Lett. Nuovo Cimento\ {\bf #1}\ }
\newcommand{\NC}[1]{Nuovo Cimento\ {\bf #1}\ }

\newcommand{\PREP}[1]{Phys. Reports \ {\bf #1} \ }
\newcommand{\RMP}[1]{Rev. Mod. Phys. \ {\bf #1} \ }
\newcommand{\IJMP}[1]{Int. J. Mod. Phys. \ {\bf #1} \ }
\newcommand{\half}{\mbox{\scriptsize $\frac{1}{2}$}}

\begin{document}

\newtheorem{thm}{Theorem}
\newtheorem{prop}{Proposition}
\newtheorem{defn}{Definition}
\newtheorem{lemma}{Lemma} \newpage
\pagenumbering{arabic} \begin{flushleft} G\"{o}teborg\\ ITP 92 - 56\\

 December 1992 \end{flushleft} \vspace{1.5cm} \begin{center} {\huge
An Algorithm for Computing Four-Ramond Vertices at Arbitrary Level}\\[1cm]
{\large Niclas
Engberg\footnote{TFENE@FY.CHALMERS.SE}\\ Bengt E.W
Nilsson\footnote{TFEBN@FY.CHALMERS.SE}\\Per
Sundell\footnote{TFEPSU@FY.CHALMERS.SE}}\\[1cm] {\sl Institute of
Theoretical Physics\\Chalmers University of Technology\\ and University of
G\"{o}teborg\\ S-412 96 G\"{o}teborg, Sweden} \end{center} \vspace{1cm}
\begin{abstract} We perform the sewing of two (dual) Ramond reggeon
vertices and derive an algorithm by means of which the so obtained
four-Ramond reggeon vertex may be explicitly computed at arbitrary
oscillator (mass) level. A closed form of the four-vertex is then deduced
on the basis of a comparison to all terms obtained by sewing that contain
only level zero and one oscillators. Results are presented for both complex
fermions and for the previously studied case of real fermions.
\end{abstract} \newpage

 \section{Introduction}

 Out of the large body of explicitly known expressions for bosonic and
fermionic string scattering amplitudes only a small subset is understood
from the point of view of operator sewing. Using reggeonic operator
vertices it is relatively straightforward to derive such amplitudes for any
genus, $g$, and number of external legs, $N$, provided only fields with
periodic (untwisted) boundary conditions in the complex plane are
considered [1-7].  As soon as one incorporates twisted fields, either
bosons or fermions, e.g. the Ramond sector of the NSR string, one
encounters severe technical problems. These problems are obvious already in
the early literature on this subject [8-18] (see also [19-21]), and have
not yet been overcome. Although there exist in many specific cases means of
getting around these obstacles, e.g. BPZ techniques \cite{BPZ},
bosonization \cite{Kni1}, overlap conditions \cite{NW88a,FW2}, path
integrals \cite{DP1,Mand1}, other operator formalisms \cite{AGMV2} etc, it
would still be of great value to find a way to derive string vertices as
well as conformal field theory (CFT) correlation functions directly by
sewing of some basic three-vertices [29-32,3], or, which is equivalent but
sometimes more convenient, dual vertices [33-35,7].

 In the case of the NSR string the problems eluded to above set in already
when trying to compute one of the most fundamental vertices namely the
four-Ramond reggeon vertex, i.e. the vertex describing the scattering of
four fermions from the spacetime target space point of view. This vertex,
usually discussed in terms of real world sheet fermions, has been the main
object of interest in a large number of previous works (see e.g.
\cite{CO,Cor,SW1,CGOS,SW3,BCO}) and will only be commented upon towards the
end of this paper. Instead we concentrate here on the slightly simpler case
based on complex fermions for which the results presented are quite
detailed. The reason for doing this is that the main technical problems to
be addressed here are basically independent of whether we use real or
complex fields, and that most of our results are somewhat easier to derive
and to present in terms of complex fields.  It deserves to be emphasized,
however, that the same kind of results are easily extracted also for the
standard NSR case based on real fermions and we collect the formulae for
this case in section 5.

 Another reason for being interested in the case of complex fermions is the
possibility of normal ordering also the zero modes of the twisted fields
which is not possible in the real case. In fact, the starting point for our
investigations is an expression for the dual Ramond reggeon vertex which is
independent of the normal ordering chosen and of whether we use complex or
real fermion fields.  The sewing computations as well as the complications
that arise when trying to put the results into closed form are essentially
the same as in the case of the standard four-Ramond vertex, and our main
goal is to extend significantly on previous sewing results in a simpler
setting and to propose closed answers for both complex and real
four-reggeon vertices.  The form of these closed answers are inspired by
three-vertices previously discussed by LeClair in \cite{LeCl2}. A
generalization of these latter vertices to an arbitrary number of $NS$ and
$R$ legs was subsequently suggested in \cite{DHMR}, but was only compared
to the sewed operator expression for the one non-zero term containing only
zero modes.  However, with our new results we can compute the four-vertex
at any oscillator (mass) level and are thus in effect able to prove that
the proposed answer is the only possible one. (A detailed comparison is
carried out below for all terms containing only level one and/or zero
oscillators.) A step by step derivation of the proposed closed form of the
four-vertex is a more difficult task that we hope to come back to in a
future publication. We also hope to be able to generalize this use of
complex fields to other CFT's and thereby extend this formalism to cases
where sewing has not yet been attempted.

 The paper is organized as follows. In section 2 we set up the formalism
and present a form of the dual twisted (Ramond) reggeon vertex from which
one can deduce both the real and the complex vertices of interest to this
paper. The explicit form of these vertices are given at the end of that
section. The complex one is then used, in section 3, to sew together two
dual twisted reggeon vertices. The so obtained four-reggeon vertex is
expressed in terms of an infinite set of quantities constructed from
infinite dimensional matrices and vectors. These quantities are then
studied and an algorithm presented by means of which they may all be
computed. Explicit answers are derived for a large number of them which are
then used, in section 4, to deduce the closed form of the twisted
four-reggeon vertex.  Formulae extending our conclusions to three other
closely related vertices for complex fermions as well as to the standard
four-Ramond reggeon vertex (derived using real fermions) together with some
other relevant facts are collected in section 5. Some further comments and
conclusions are given in section 6.

 \section{Background and notation}

 All previous sewing results in connection with reggeon four-vertices for
external twisted fermionic fields have been obtained using the real
fermions ($\;''\;\hat{}\;''\;$ denotes operator valued quantities)
\eqs{\hat{\psi}^{\m}_{NS}(z)=i\sum_{r\in {\bf
Z}+{1\/2}}\hatpsi^{\m}_{r}\;z^{-r-{1\/2}}\;\;\;\;\;,\;\;\;\m=1,...,D\\
\hat{\psi}^{\m}_{R}(z)=i\sum_{m\in {\bf
Z}}\hatpsi^{\m}_{m}\;z^{-m-{1\/2}}\;\;\;\;\;,\;\;\;\m=1,...,D } whose
Laurent modes satisfy the hermiticity conditions
\eqs{(\hatpsi^{\m}_{r})^{\dagger}
=\hatpsi^{\m}_{-r},\;\;\;\;(\hatpsi^{\m}_{m})^{\dagger}=\hatpsi^{\m}_{-m} }
Our notation is such that indices $r,s$ always take values in the set ${\bf
Z}+{1\/2}$, while $m,n$ always refer to integers.  The vacua in the two
sectors, usually referred to as Neveu-Schwarz ($NS$) and Ramond ($R$),
respectively, are defined by
\eqs{\hat{\psi}^{\m}_r\ket{0}{}=0,\;\;\;\;\hat{\psi}^{\m}_n\ket{\a}{}=0;
\;\;\;\;r\geq\half,\;\;\;n\geq1 } where $\ket{\a}{}$ is a set of degenerate
states carrying a representation of the Clifford algebra satisfied by the
Ramond zero modes $\hatpsi_{0}^{\m}$.  To define the systems of main interest
here we assume $D$ to be an even integer, say $D=2d$, and combine pairs of
the real fields into complex fields as follows
\eqs{\hat{\psi}^{a}_{NS}(z)&=&{1\/\sqrt{2}}\br{\hatpsi^{a}_{NS}(z)+
i\hatpsi_{NS}^{a+d}(z)},
\;\;\;\;\hat{\bar{\psi}}^{a}_{NS}(z)={1\/\sqrt{2}}\br{\hatpsi^{a}_{NS}(z)-
i\hatpsi^{a+d}_{NS}(z)}\;\;\;,
\\
\hat{\psi}^{a}_{R}(z)&=&{1\/\sqrt{2}}\br{\hatpsi^{a}_{R}(z)+
i\hatpsi^{a+d}_{R}(z)},
\;\;\;\;\hat{\bar{\psi}}^{a}_{R}(z)={1\/\sqrt{2}}\br{\hatpsi_{R}^{a}(z)-
i\hatpsi_{R}^{a+d}(z)}\;\;,
} where the index $a$ runs over values $1,...,d$.  These complex fields are
expanded as \eqs{\hatpsi^{a}_{NS}(z)&=&i\sum_{r\in {\bf
Z}+{1\/2}}\hatpsi^{a}_{r}z^{-r-{1\/2}},
\;\;\;\;\hat{\bar{\psi}}^{a}_{NS}(z)=i\sum_{r\in{\bf
Z}+{1\/2}}\hat{\bar{\psi}}^{a}_{r}z^{-r-{1\/2}}\lb
\hatpsi^{a}_{R}(z)&=&i\sum_{m\in {\bf
Z}}\hatpsi^{a}_{m}z^{-m-{1\/2}},\;\;\;\;
\hat{\bar{\psi}}^{a}_{R}(z)=i\sum_{n\in {\bf
Z}}\hat{\bar{\psi}}^{a}_{n}z^{-n-{1\/2}} } where the respective modes
satisfy the hermiticity properties
\eqs{(\hatpsi^{a}_{r})^{\dagger}=\hat{\bar{\psi}}^{a}_{-r},\;\;
(\hat{\bar{\psi}}^{a}_{r})^{\dagger}=\hatpsi^{a}_{-r},\;\;
(\hatpsi^{a}_{m})^{\dagger}=\hat{\bar{\psi}}^{a}_{-m},\;\;
(\hat{\bar{\psi}}^{a}_{m})^{\dagger}=\hatpsi^{a}_{-m} } It is now
convenient to summarize all non-zero oscillator commutation relations as
follows: \eqs{\left\{ \hat{\psi}^{a(+)}_{NS}(z),\hat{
\bar{\psi}}^{b(-)}_{NS}(w) \right\}= \left\{
\hat{\bar{\psi}}^{a(+)}_{NS}(z),\hat{ \psi}^{b(-)}_{NS}(w) \right\}=
-{\d^{ab}\/z-w},\;\;\;|z|>|w| \eqn{NSprop} } in the $NS$ sector, \eqs{
\left\{ \hat{\psi}^{a(+)}_{R}(z),\hat{ \bar{\psi}}^{b(-)}_{R}(w) \right\}=
-{\sqrt{z\/w}}{\d^{ab}\/z-w},\;\;\;\; \left\{
\hat{\bar{\psi}}^{a(+)}_{R}(z),\hat{ \psi}^{b(-)}_{R}(w) \right\}=
-{\sqrt{w\/z}}{\d^{ab}\/z-w},\;|z|>|w| \eqn{Rprop} } in the $R$ sector.
Here we have defined creation $(_)$ and annihilation $(+)$
fields by the following expansions:
\eqs{\hatpsi_{NS}^{a(+)}(z)=i\sum_{r=\half}^{\infty}\hatpsi^{a}_{r}
z^{-r-{1\/2}},\;\;
\hatpsi_{NS}^{a(-)}(z)=i\sum_{r=\half}^{\infty}\hatpsi^{a}_{-r}
z^{r-{1\/2}},\;\;\lb
\hat{\bar{\psi}}_{NS}^{a(+)}(z)=i\sum_{r=\half}^{\infty}
\hat{\bar{\psi}}^{a}_{r}z^{-r-{1\/2}},\;\;
\hat{\bar{\psi}}_{NS}^{a(-)}(z)=i\sum_{r=\half}^{\infty}
\hat{\bar{\psi}}^{a}_{-r}z^{r-{1\/2}},
\eqn{NS+-} \\
\hatpsi_{R}^{a(+)}(z)=i\sum_{n=0}^{\infty}\hatpsi^{a}_{n}
z^{-n-{1\/2}},\;\;
\hatpsi_{R}^{a(-)}(z)=i\sum_{n=1}^{\infty}\hatpsi^{a}_{-n}
z^{n-{1\/2}},\;\;\lb
\hat{\bar{\psi}}_{R}^{a(+)}(z)=i\sum_{n=1}^{\infty}\hat{\bar{\psi}}^{a}_{n}
z^{-n-{1\/2}},\;\;
\hat{\bar{\psi}}_{R}^{a(-)}(z)=i\sum_{n=0}^{\infty}\hat{\bar{\psi}}^{a}_{-n}
z^{n-{1\/2}}
\eqn{R+-} } which explains the different square root factors appearing on
the right hand sides of the $R$ commutators in eq. \refbr{Rprop} as due to
the $R$ zero mode.  The ket vacua of the complex fermions
can thus be defined through the equations
\eqs{\hat{\psi}^{a(+)}_{NS}(z)\ket{0}{}&=&\hat{\bar{\psi}}^{a(+)}_{NS}(z)
\ket{0}{}={}0
\\
\hat{\psi}^{a(+)}_{R}(z)\ket{0}{}&=&\hat{\bar{\psi}}^{a(+)}_{R}(z)
\ket{0}{}={}0
} while for the bra vacua the corresponding equations follow from hermitian
conjugation of the oscillators and $\bra{0}{}= (\ket{0}{})^{\dagger}$. Note
that notationally we do not distinguish between the $NS$ vacuum and the
Clifford representation vacuum in the complex $R$ sector.

  By introducing two $NS$ type normal ordering fields $\hat{\psi}_1$ and
$\hat{\psi}_2$ together with one $NS$ type auxiliary field
$\hat{\psi}_{aux}$ the basic (dual) vertex can be written, in either the
real or complex $NS$ and $R$ sectors, in the following forms:
\eqs{\hat{W}_{ext}(V)&=&\bra{0}{1}\bra{0}{2}
e^{\oint_{C}dz(\hat{\psi}^{\m}_{aux}+i\hat{\psi}^{\m}_{1})(z)
(\hat{\psi}^{\m}_{ext}+i\hat{\psi}^{\m}_{2})^{V^{-1}}(z)}
\ket{0}{2} \ket{0}{1}=\eqn{realform}\\ &=&\bra{0}{1}\bra{0}{2}
e^{\oint_{C}dz\left((\hat{\psi}_{aux}^{a}+i\hat{\psi}_{1}^{a})(z)
(\hat{\bar{\psi}}_{ext}^{a}+
i\hat{\bar{\psi}}_{2}^{a})^{V^{-1}}(z)+
(\hat{\bar{\psi}}_{aux}^{a}+i\hat{\bar{\psi}}_{1}^{a})(z)
(\hat{\psi}_{ext}^{a}+
i\hat{\psi}_{2}^{a})^{V^{-1}}(z)\right) }\ket{0}{2} \ket{0}{1}
\eqn{complexform}}
where $V$ denotes a projective transformation and where the contour $C$
encircles the poles of the $NS$ fields indexed $aux$ and $1$, and where the
index $ext$ stands for either $NS$ or $R$. Furthermore, the explicit normal
ordering of the auxiliary and external fields is induced by performing the
correlation functions for the two normal ordering fields as indicated in
the equation for the vertex. Note that the complex form \refbr{complexform}
is only defined when $D$ is even, and in that case
\refbr{realform} equals \refbr{complexform}.

 It is convenient to present the vertex in this form for several reasons.
One point is connected to the transformation properties of the vertex. When
computing e.g. vertices at tree level one has to ensure that the emission
points of the external legs are distinct. Multilegged vertices are derived
by computing a correlation in the auxiliary fields involving dual vertices;
one example is the calculation performed in section three.  Now, due to the
fact that the normal ordering fields are always (i.e.
irrespective of the choice of external legs) of $NS$ type, the necessary
transportation of the vertices to different emission points in the complex
plane can be accomplished rather easily. By acting with
$\hat{\g}_{aux}(V)$, the operator generating the projective transformation
$V$, from the left and its inverse from the right, and observing that any
projective transformation can be extracted from the $NS$ vacua of the
normal ordering field $\hat{\psi}_1$, the above form of the vertex is an
immediate consequence of the transformation law for the $NS$ fields
themselves \eq{\hat{\g}_{NS}(V) \hat{\psi}_{NS}(z)
\hat{\g}^{-1}_{NS}(V)=\sqrt{V'(z)}\hat{\psi}_{NS}(V(z))=\hat{\psi}^V_{NS}(z)
 }
This equation also defines what we mean by $\hat{\psi}^V_{NS}$.  As
explained in \cite{ENS1} the extra fields must all be of $NS$ type to
ensure that the vertex behaves correctly under these transformations.

 A second reason for introducing the normal ordering fields is that it
clarifies how the
different explicit normal ordering prescriptions that are commonly used for
the $R$ zero modes are related
to each other. In general
different options arise because when splitting the fields
into two or three parts (e.g. creation, zero modes, and annihilation parts)
this can be done in any way one chooses as long as, when the
Baker-Hausdorff (BH) formula is used to put the creation and annihilation
fields into separate exponentials, one finds that in the commutator term
the residue at $z=w$ from the propagator of the
$aux$ ($ext$) field cancels against the
corresponding residue from the propagator of the normal ordering field
${\hat{\psi}}_1$ (${\hat{\psi}}_2$).  Note that one may leave a finite
number of modes unordered and still have a well defined vertex.  In the NSR
string constructed from real fermions this is normally done for the zero
modes of the $R$ fields while all other modes are normal ordered the usual
way. We denote this particular normal ordering by
${}^{\times}\!\!\!\!_{\times}\; ^{\times}\!\!\!\!_{\times}$. In the case of
complex fermions one usually normal orders all the modes of the $R$ fields
and for this normal ordering we will use the notation $:\; :$.  Both real
and complex $NS$ fields are normal ordered in all their modes as implied
here by both the above normal ordering symbols.

 For the dual $NS$ vertex both the $aux$ field and the $ext$ field are of
$NS$ type. Hence, when computing the correlations functions appearing in
the above forms of the dual vertex, using the split in \refbr{NS+-} (for
complex fermions and the corresponding split for real fermions), one
finds that the BH commutator term cancels completely. Doing the same
for for vertices with an external twisted field split as in \refbr{R+-}
(for complex fermions and the corresponding split for real fermions)
one is however left with a
remaining term. This term is the actual origin of all the complications we
have to face in these twisted cases. To see this explicitly we use the $R$
and $NS$ commutators given in \refbr{Rprop} and \refbr{NSprop} to compute,
in the Ramond case, the total propagator in the BH commutator term. One
finds that it does not vanish but is instead proportional to
\eq{{-\sqrt{{w\/z}}+1\/z-w}\;\;\;or\;\;\;{-\sqrt{{z\/w}}+1\/z-w}\;\;\;, }
which indeed have zero residues. Thus by expressing the projective
transformation $V(z)$ in terms of the emission points $z_1,z_2$ of the
external legs (and a parameter $a$) through its inverse
$V^{-1}(z)=a{z-z_1\/z-z_2}$, and eliminating the normal ordering fields,
the dual reggeon vertex becomes \eqs{\hat{W}_{R}(V)&=&:e^{\oint_{C}
dz(\hatpsi^{a}_{aux}(z)\hat{\bar{\psi}}_{R}^{aV^{-1}}
(z)+\hat{\bar{\psi}}^{a}_{aux}(z)\hatpsi_{R}^{aV^{-1}}(z))} \lb
&&e^{-{1\/2}\oint_{C}dz\oint_{C}dw(\hatpsi^{a}_{aux}(z)M(z_{1},z_{2};z,w)
\hat{\bar{\psi}}^{a}_{aux}(w)+
\hat{\bar{\psi}}^{a}_{aux}(z)\bar{M}(z_{1},z_{2};z,w)\hatpsi^{a}_{aux}(w))}
: \;\;\;=\eqn{dualcomplexReqn}\\
&=&^{\times}\!\!\!\!_{\times}e^{\oint_{C}dz\hatpsi^{\m}_{aux}(z)
\hatpsi_{R}^{\m
V^{-1}}(z)}\lb && e^{-{1\/2}\oint_{C}dz\oint_{C}dw
\hatpsi_{aux}^{\m}(z)H(z_{1},z_{2};z,w)\hatpsi_{aux}^{\m}(z)}\;
{}^{\times}\!\!\!\!_{\times}\eqn{dualrealR} } when $ext$ equals $R$, and
\eqs{\hat{W}_{NS}(V)=&=&:
e^{\oint_{C}dz(\hatpsi_{aux}^{a}\hat{\bar{\psi}}^{aV^{-1}}_{NS}(z)+
\hat{\bar{\psi}}^{a}_{aux}(z)\hatpsi^{aV^{-1}}_{NS}(z))}{}:\;\;\;=\\ &=& :
e^{\oint_{C}dz\hatpsi^{\m}_{aux}(z)\hatpsi^{\m V^{-1}}_{NS}(z)}{}:} when
$ext$ equals $NS$.  Here the contours $C$ always encircle the poles of the
$\psi_{aux}$ fields, and : : and
$^{\times}\!\!\!\!_{\times}\;^{\times}\!\!\!\!_{\times}$ refer to the
normal ordering of both $aux$ and $ext$ fields.  In the four-vertices
\refbr{dualcomplexReqn} and \refbr{dualrealR} we have also defined
\eqs{M(z_1,z_2;z,w)&=&{-\sqrt{{w-z_{1}\/w-z_{2}}{z-z_{2}\/z-z_{1}}}+1\/z-w}
\eqn{M} \\
\bar{M}(z_1,z_2;z,w)&=&{-\sqrt{{z-z_{1}\/z-z_{2}}{w-z_{2}\/w-z_{1}}}+1\/z-w}
\eqn{Mbar} \\ H(z_1,z_2;z,w)&=&{1\/2}(M(z_1,z_2;z,w)+\bar{M}(z_1,z_2;z,w))
\eqn{H} } Note that the parameter $a$ cancels in $M$ and $\bar{M}$ and
therefore appears in the four-vertices \refbr{dualcomplexReqn} and
\refbr{dualrealR} only in the transported $R$ fields. Let us once again
remind the reader that different normal orders are taken into
account in \refbr{dualcomplexReqn} and \refbr{dualrealR} and therefore
these to forms look
different while in fact being equal (for even $D$) as they are both
given by the same correlation functions \refbr{realform} and
\refbr{complexform} above.

In the next section we will use the following matrix form of
the dual reggeon vertices in the $R$ sector:
\eqs{\hat{W}_{R}(V)&=&
:e^{-\hat{\barpsi}_{-r}^{a}U_{rm}^{(-)}(V)\hatpsi_{m}^{R,a}-
\hat{\barpsi}_{r}^{a}U_{rm}^{(+)}(V)\hatpsi_{m}^{R,a}-
\hatpsi_{-r}^{a}\barU_{rm}^{(-)}(V)\hat{\barpsi}_{m}^{R,a}-
\hatpsi_{r}^{a}\barU_{rm}^{(+)}(V)\hat{\barpsi}_{m}^{R,a}}\lb
&{}&
e^{\half(\hat{\barpsi}_{-r}^{a}M^{(--)}_{rs}(V)\hatpsi_{-s}^{a}+
\hat{\barpsi}_{-r}^{a}M^{(-+)}_{rs}(V)\hatpsi_{s}^{a}+
\hat{\barpsi}_{r}^{a}M^{(+-)}_{rs}(V)\hatpsi_{-s}^{a}+
\hat{\barpsi}_{r}^{a}M^{(++)}_{rs}(V)\hatpsi_{s}^{a})}\lb
&{}&
e^{\half(\hat{\psi}_{-r}^{a}\barM^{(--)}_{rs}(V)\hat{\barpsi}_{-s}^{a}+
\hat{\psi}_{-r}^{a}
\barM^{(-+)}_{rs}(V)\hat{\barpsi}_{s}^{a}+\hat{\psi}_{r}^{a}
\barM^{(+-)}_{rs}(V)\hat{\barpsi}_{-s}^{a}+
\hat{\psi}_{r}^{a}\barM^{(++)}_{rs}(V)\hat{\barpsi}_{s}^{a})}:\;\;\;=
\eqn{dualcomplexRvertexonmatrixform}\\ &=& {}^{\times}\!\!\!\!_{\times}
e^{-\hatpsi_{-r}^{\m}U^{(-)}_{rm}(V)\hatpsi^{R,\m}_{m}-
\hatpsi_{r}^{\m}U^{(+)}_{rm}(V)\hatpsi^{R,\m}_{m}}\lb &{}&
e^{\half(\hatpsi^{\m}_{-r}H^{(--)}_{rs}(V)\hatpsi^{\m}_{-s}+
\hatpsi^{\m}_{-r}H^{(-+)}_{rs}(V)\hatpsi^{\m}_{s}+
\hatpsi^{\m}_{r}H^{(+-)}_{rs}(V)\hatpsi^{\m}_{-s}+
\hatpsi^{\m}_{r}H^{(++)}_{rs}(V)\hatpsi^{\m}_{s})}
{}^{\times}\!\!\!\!_{\times}\eqn{dualrealRvertexonmatrixform}}
where $\hatpsi_{r}$ and $\hatpsi^{R}_{m}$
are the auxiliary and Ramond  Laurent modes respectively ,
and where we have defined the
following matrices \eqs{M^{(\pm\pm)}_{rs}(V)&=&
\oint_{C}dz\oint_{C}dwz^{\mp
r-{1\/2}}M(z_{1},z_{2};z,w)w^{\mp s -{1\/2}}\\
\barM^{(\pm\pm)}_{rs}(V)&=&\oint_{C}dz\oint_{C}dwz^{\mp
r-{1\/2}}\barM(z_{1},z_{2};z,w)w^{\mp s -{1\/2}}\\
H^{(\pm\pm)}_{rs}(V)&=&{1\/2}(M^{(\pm\pm)}_{rs}(V)+
\barM^{(\pm\pm)}_{rs}(V))\\
U_{rm}^{(\pm)}(V)&=&\bar{U}_{rm}^{\pm}(V)=\oint_{C}dzz^{\mp
r-{1\/2}}\sqrt{V^{-1'}(z)}(V^{-1}(z))^{-m-{1\/2}}\eqn{defofumatrix}} with
$C$ encircling zero and infinity.  Note that $m,n$ are summed over all
integers, while $r,s$ are summed over ${1\/2},{3\/2},...$.

 In the next section we give the corresponding expression for the sewed
complex four-Ramond reggeon
vertex and proceed to extract the quantities that we need to compute
in order to deduce the closed form (see section 4) of the vertex.
In section 5 we will give the equivalent expressions for real
Ramond fermions.

 \section{The sewed complex four-Ramond reggeon vertex}

 The reggeon vertex for four external twisted
fermionic (Ramond)
complex fields is obtained by computing the auxiliary correlation function
\eq{\hat{W}_{R_1R_2}(V_1,V_2)=\bra{0}{aux}\hat{W}_{R_1}(V_1)
\hat{W}_{R_2}(V_2)\ket{0}{aux}
} where the dual reggeon vertices are given in \refbr{dualcomplexReqn}
(note that each index $R$ refers to $two$ external legs).

 Before we proceed we need to specify the projective transformations used
in each one of the two vertices.  Let us choose the emission points
$V(\infty)$ and $V(0)$ to be $\infty$ and ${-1\/\l}$, respectively, for
vertex 1, and $-\l$ and 0, respectively, for vertex 2 with $|\l|<1$. This
can be implemented by the following choice of projective transformations
\eqs{V_{1}(z)=z-{1\/\l},\;\;\;\;\;V_{2}(z)=-{z\/{z\/\l}-1}
\eqn{thechoiceofvs} } whose inverses are
\eqs{V_{1}^{-1}(z)=z+{1\/\l},\;\;\;\;V_{2}^{-1}(z)={z\/{z\/\l}+1} } This
choice implies that the matrix form of the four-vertex becomes (we suppress
the spacetime indices henceforth) \eqs{ \hatW_{R_{1}R_{2}}(\l)= :
\bra{0}{aux}
e^{-\hat{\bar{\psi}}_{r}U_{R_{1},r}^{(+)}-\hatpsi_{r}\barU_{R_{1},r}^{(+)}}
e^{{1\/2}\hatpsi_{r}M_{rs}^{(++)}(V_{1})\hat{\bar{\psi}}_{s}+
{1\/2}\hat{\bar{\psi}}_{r}\bar{M}_{rs}^{(++)}(V_{1})\hatpsi_{s}}\lb
e^{-\hat{\bar{\psi}}_{-r}U_{R_{2},r}^{(-)}-
\hatpsi_{-r}\barU_{R_{2},r}^{(-)}}
e^{{1\/2}\hatpsi_{-r}M_{rs}^{(--)}(V_{2})\hat{\bar{\psi}}_{-s}+
{1\/2}\hat{\bar{\psi}}_{-r}\bar{M}_{rs}^{(--)}(V_{2})\hatpsi_{-s}}
\ket{0}{aux} : } where the normal ordering now refers only to the $R$
oscillators in the $U$'s (see below) and where $\hat{\psi}_r$ are the
auxiliary modes. Also, \eq{ \begin{array}{ll}
M^{(++)}(V_{1})=M,\;\;\;\;\;\;\;\;&\bar{M}^{(++)}(V_{1})=-M^T\\
M^{(--)}(V_{2})=M^T,\;\;\;\;\;\;&\bar{M}^{(--)}(V_{2})=-M
\end{array}\eqn{resultofchoiceofvs} } where the matrix $M$ is defined by
\cite{CO} \eqs{M_{rs}(\l)=\oint_{0}dz\oint_{0}dwz^{-r-{1\/2}}w^{-s-{1\/2}}
{1-\sqrt{{w+{1\/\l}\/z+{1\/\l}}}\/z-w} ={r\/r+s}\left( \begin{array}{c}
-{1\/2} \\ r-{1\/2}\end{array}\right) \left(\begin{array}{c} -{1\/2} \\
s-{1\/2}\end{array}\right)\l^{r+s} } An important consequence of the
particular choice of emission points made in \refbr{thechoiceofvs} is the
very simple relation between $M^{(++)}(V_{1})$ and $M^{(--)}(V_{2})$; see
\refbr{resultofchoiceofvs} above (or \refbr{MT} below), and as will be
obvious later much of the analysis carried out in this paper relies heavily
on this simple relation. We will return to the question of using arbitrary
projective transformations in section 5. The other quantities in the vertex
to be defined are the vectors $U^{(\pm)}_r$. They are given by
\eqs{\barU^{(+)}_{R_{1},r}=\barU_{rm}^{(+)}(V_{1})\hat{\bar{\psi}}_{m}^{R_{1}}&,
&\;\;\;\;U_{R_{1},r}^{(+)}=U_{rm}^{(+)}(V_{1})\hatpsi_{m}^{R_{1}}\lb
\barU_{R_{2},r}^{(-)}=\barU_{rm}^{(-)}(V_{2})\hat{\bar{\psi}}_{m}^{R_{2}}&,
&\;\;\;\;U_{R_{2},r}^{(-)}=U_{rm}^{(-)}(V_{2})\hatpsi_{m}^{R_{2}} }
For the above choices for $V_1$ and $V_2$ the matrices
$U^{(\pm)}_{rm}(V_{i})$ become
\eqs{ U_{rm}^{(+)}(V_{1})=\sqrt{2}v_{r}^{(m)}\lb
U_{rm}^{(-)}(V_{2})=-\sqrt{2}v_{r}^{(-m)} }
where we have defined
\eq{v^{(m)}_r={1\/\sqrt{2}}\left(\begin{array}{c} -m-{1\/2} \\ r-{1\/2}
\end{array}\right) \l^{m+r}}

Then by inserting coherent states between the reggeons 1 and 2, and
performing the so obtained infinite dimensional integral one finds, for d
complex fields, that
\eqs{\hatW_{R_{1}R_{2}}(\l)=\left[det(1+MM^{T})\right]^{d}:
\exp{\left(\barU_{R_{2}}^{T(-)}{1\/1+M^{T}M}
U_{R_{1}}^{(+)}+U_{R_{2}}^{T(-)}{1\/1+MM^{T}}\barU_{R_{1}}^{(+)}
\right.} \lb
\left. + U_{R_{2}}^{T(-)}M{1\/1+M^{T}M}\barU_{R_{2}}^{(-)}-
\barU_{R_{1}}^{T(+)}M{1\/1+M^{T}M}U_{R_{1}}^{(+)}\right) : \eqn{matrixW} }

We are now in the position where we can enumerate the quantities needed for
the comparison of the above results to the proposed closed answer in the
next section, and establish its correctness. Having already defined the
infinite dimensional vectors $v^{(m)}$ the relevant quantities are
\eq{X^{(m,n)}=v^{(m)T}\xi^{(n)},\;\;\;\;\;\;
    Y^{(m,n)}=v^{(m)T}\eta^{(n)}, } where
\eqs{\xi^{(m)}={1\/1+M} v^{(m)}, \;\;\;\;\;\;
     \eta^{(m)}={1\/1-M} v^{(m)},} which generalize quantities introduced
in the early 1970's [8-17] (corresponding to $m=n=0$ of the above). In
these references one may also find some of the tricks used below to
manipulate these infinite dimensional matrices and vectors.  The fact that
all terms in the exponent of $\hat{W}_{R_1R_2}$ can be expressed in terms
of $X^{(m,n)}$ and $Y^{(m,n)}$ can be seen as follows. First note that,
writing $v$ instead of $v^{(0)}$,
\eq{M_{rs}=H_{rs}+v_r v_s}
where $H$ is the antisymmetric part of $M$, and hence
\eq{M^T=-M+2vv^T\eqn{MT}.}
Then the extremely useful matrix identity
\eq{A^{-1}(A-B)B^{-1}=B^{-1}-A^{-1}\eqn{ABeq}}
with $A=1-M^2$, $B=1+M^T M = 1-M^2 +2vv^TM$ gives
\eq{{1\/1+M^TM}={1\/1-M^2}-2{1\/1-M^2}vv^TM{1\/1+M^TM}\eqn{MTM}}
Multiplying this equation by $v^TM$ from the left produces an expression
for $v^TM{1\/1+M^TM}$ that can be inserted back into the last term on the
right hand side of the last equation. Doing so, one finds
\eq{{1\/1+M^TM}={1\/1-M^2}-2{{1\/1-M^2}vv^T{M\/1-M^2}\/1+2v^T{M\/1-M^2}v}
\eqn{MTMM}}
Finally, the identities
\eqs{{1\/1-M^2}={1\/2}\left({1\/1-M}+{1\/1+M}\right),\;\;\;\;\;\;
     {M\/1-M^2}={1\/2}\left({1\/1-M}-{1\/1+M}\right)} proves the assertion
stated above.

Although we will later present explicit formulae only for those terms in
the exponent of $\hat{W}_{R_1R_2}$ that are expressible in terms of
$X^{(m,n)}$ and $Y^{(m,n)}$ with $m,n=0,\pm1$, we emphasize here that it is
possible to derive answers for all integers $m,n$. To explain how this can
be done we divide the quantities $X^{(m,n)}$ and $Y^{(m,n)}$ into two groups;
those with both $m$ and $n$ non-negative or with either $m$ or $n$
negative, and those with $m$ and $n$ both negative. The reason for this
particular division will become clear below.

To deal with the first group we proceed as follows. As explained in detail
elsewhere \cite{SW1,CGOS,NS1}, it is possible to show that
\eq{\xi_r={1\/\sqrt{2}} \oint_{C}{dt\/2\pi it}(w_{\l}(t))^r \eqn{xi}}
where
\eq{w_{\l}(t)={t-{(1+\l)^2\/4\l}\/t(1-t)}\eqn{w}}
and where the contour $C$ encircles the branch cut between t=0 and t=1,
solves the equation
\eq{(1+M)\xi=v\eqn{xiv}}
and hence to conclude that $\xi$ is identical to what we above denoted as
$\xi^{(0)}$.  To prove that $\xi$ satisfies \refbr{xiv}, one makes use of
among other things the following change of variables
\eq{t\mapsto{(1+\l)^2\/4\l t}}
which has the effect that
\eq{w_{\l}(t)\mapsto{1\/w_{\l}(t)}}
and that $C$ after the change encircles the cut between $t={(1+\l)^2\/4\l}$
and $t=\infty$. Flipping the contour back to its original place gives the
result that $\xi_r$ can also be written
\eq{\xi_r={1\/\sqrt{2}} \oint_{C}{dt\/2\pi it}(w_{\l}(t))^{-r}}
(note the sign change in the exponent). Using either expression for
$\xi_r$, one can easily do the sum in
$v^T\xi=\sum_{r={1\/2}}^{\infty}v_r\xi_r$, and then perform a
straightforward contour integral around two poles to find \cite{CGOS}
\eq{X^{(0,0)}=v^T\xi={1\/2}(1-\sqrt{{1-\l\/1+\l}})\eqn{X00}}
The corresponding formula for $\eta=\eta^{(0)}$ can be found from the
$\l$-derivative of $\xi$. To get this relation, we define the matrix $R$ by
\eq{R_{rs}=r\d_{rs}\eqn{defofRmatrix}}
and note that
\eq{MR=RM^T\eqn{MR}}
{}From the definitions of $M$ and $v$ it follows that
\eq{\l{\partial\/\partial\l}M=2Rvv^T,\;\;\;\;\;\;\l{\partial\/\partial\l}v=
Rv\eqn{derMv}}
which implies that
\eq{\l{\partial\/\partial\l}{1\/1+M}=
-{1\/1+M}(\l{\partial\/\partial\l}M){1\/1+M}=-2{1\/1+M}Rvv^T{1\/1+M}\eqn{derM}}
Using this equation when computing the $\l$-derivative of the equation
$(1+M)\xi=v$ gives
\eq{\l{\partial\/\partial\l}\xi=R\tilde{\xi}(1-2v^T\xi)}
where we have defined
\eq{\tilde{\xi}={1\/1+M^T}v}
Next we define $\tilde{\eta}$ in an analogous fashion, and derive formulae
relating them to their untilded counterparts by setting $A=1\pm M^T$ and
$B=1\mp M$ in the matrix equation \refbr{ABeq} and multiplying it by $v$
from the right. The result is
\eq{\tilde{\xi}={\eta\/1+2v^T\eta},\;\;\;\;\;\;\;\;\tilde{\eta}=
{\xi\/1-2v^T\xi}\eqn{tilde}}
for the upper and lower sign, respectively. Hence
\eq{\l{\partial\/\partial\l}\xi=R\eta{1-2v^T\xi\/1+2v^T\eta}\eqn{derxi}}
Multiplying also from the left with $v^T$ gives
\eq{v^T\xi={v^T\eta\/1+2v^T\eta},\;\;\;\;\;\;\; v^T\eta={v^T\xi\/1-2v^T\xi}}
Thus eq. \refbr{X00} implies that
\eq{Y^{(0,0)}=v^T\eta=-{1\/2}(1-\sqrt{{1+\l\/1-\l}})\eqn{Y00}}
and, finally, from \refbr{derxi} we obtain
\eq{\eta_r=\mp {1\/\sqrt{2}}\oint_C{dt\/2\pi it}
\left( w_{\l}(t)\right) ^{\pm r}{1\/1-{4\l t\/(1+\l)^{2}}}\eqn{eta}}
where $w_{\l}(t)$ and the contour $C$ are as defined for $\xi_r$ and where
the two cases, differing by sign, are related through the same change of
the variable $t$ as discussed above.

Note that, as previously found in \cite{CGOS},
\eq{1-2v^T\xi=(1+2v^T\eta)^{-1}}
and that these quantities correspond to what in the bosonic case is denoted
as $g^{1\/2}$ in \cite{DGM2}.  This relation is but a special case of the
general relations between $X^{(m,n)}$ and $Y^{(m,n)}$:
\eqs{Y^{(m,n)}=X^{(n,m)}+2Y^{(m,0)}X^{(n,0)}  \eqn{YXrel}  \\
     X^{(m,n)}=Y^{(n,m)}-2X^{(m,0)}Y^{(n,0)} \eqn{XYrel} } derivable by
using $A=1\mp M$ and $B=1\pm M^T$ in eq.\refbr{ABeq}. However, comparing
the explicit expressions \refbr{X00} and \refbr{Y00}
for $X^{(0,0)}$ and $Y^{(0,0)}$ above we discover a
possible direct connection between them under $\l$ going to $-\l$. By
expanding ${1\/1\pm M}$ in geometric series and comparing the power series
in $\l$ of $X^{(m,n)}$ and $Y^{(m,n)}$ we discover immediately the powerful
relation
\eq{Y^{(m,n)}(\l)=(-1)^{m+n+1}X^{(m,n)}(-\l)\eqn{XYL}}
valid for all integers $m,n$. This equation will be discussed again below.

To summarize, we have found explicit expressions for $\xi_r$ and $\eta_r$,
and from these one can derive expressions for $\tilde{\xi_r}$ and
$\tilde{\eta_r}$ as well as for $X^{(0,0)}$ and $Y^{(0,0)}$. In fact, in
the same spirit as $X^{(0,0)}$ and $Y^{(0,0)}$ are derived from $\xi_r$ and
$\eta_r$, one may also obtain closed answers for $X^{(m,0)}$ and
$Y^{(m,0)}$ for $m$ any integer. By transposing and utilizing the formulae
for $\tilde{\xi_r}$ and $\tilde{\eta_r}$ given above, also $X^{(0,m)}$ and
$Y^{(0,m)}$ are easily found.

The quantities in the first group not yet discussed are the ones with both
$m$ and $n\neq 0$. By transposing them if necessary these can always be
written $X^{(m,n)}$, $Y^{(m,n)}$ with $n$ a positive integer. We now take
advantage of the following property of the $v^{(n)}$'s
\eq{v^{(n+1)}=\l {n+R\/n+{1\/2}} v^{(n)}\eqn{vm+1}}
to step by step lower the superindex $n$ until it reaches $n=0$. The
factors $ {n+R\/n+{1\/2}}$ (recall that $R$ is the matrix defined in
\refbr{defofRmatrix}) may then be
brought through ${1\/1\pm M}$ using ${1\/1\pm M}R=R {1\/1\pm M^T}$ thereby
producing a series of terms when hitting $v^{(m)T}$. Thus all $X^{(m,n)}$,
$Y^{(m,n)}$ in the first group can be evaluated once $X^{(m,0)}$ and
$Y^{(m,0)}$ are known.

Before turning to the second group of $X^{(m,n)}$, $Y^{(m,n)}$, i.e. those
with both $m$ and $n$ negative integers, we apply the steps discussed above
to arrive at the following expressions for the quantities in the first
group (valid for the same $\l$ on the two sides of the equations)
\eq{
\begin{array}{ll}
X^{(1,1)}={3Y^{(2,0)}-2\l Y^{(1,0)}\/1+2Y^{(0,0)}} &
Y^{(1,1)}={3X^{(2,0)}-2\l X^{(1,0)}\/1-2X^{(0,0)}} \\
X^{(0,1)}={Y^{(1,0)}\/1+2Y^{(0,0)}} & Y^{(0,1)}={X^{(1,0)}\/1-2X^{(0,0)}}
\\ X^{(-1,1)}={2\l Y^{(-1,0)}-4Y^{(0,0)}\/1+2Y^{(0,0)}} & Y^{(-1,1)}={2\l
X^{(-1,0)}-4X^{(0,0)}\/1-2X^{(0,0)}} \\
X^{(1,-1)}=Y^{(-1,1)}-{2Y^{(-1,0)}\/1+2Y^{(0,0)}}Y^{(0,1)} &
Y^{(1,-1)}=X^{(-1,1)}+{2X^{(-1,0)}\/1-2X^{(0,0)}}X^{(0,1)}\\
X^{(0,-1)}={Y^{(-1,0)} \/1+2Y^{(0,0)}} & Y^{(0,-1)}={X^{(-1,0)}
\/1-2X^{(0,0)}}
\end{array} \label{Xs2} }
where we should note that to evaluate $X^{(1,1)}$ we need $Y^{(2,0)}$ and
similarly with X and Y interchanged.

When we now turn to the second group of $X^{(m,n)}$, $Y^{(m,n)}$'s we will
resort to different methods, since using \refbr{vm+1} also here does not
seem so convenient. More specifically, for these quantities we instead
derive differential equations in $\l $. From \refbr{derMv} and \refbr{derM}
together with the relation
\eq{\l{ \partial\/\partial \l }v^{(m)}={m+{1\/2}\/ \l}v^{(m+1)}\eqn{derv}}
 we find that acting with the logarithmic $\l $-derivative on
$X^{(m,n)}=v^{(m)T}{1\/1+M}v^{(n)}$ gives
\eqs{\l {\partial\/ \partial \l} X^{(m,n)}= {1\/ \l}
\left( (m+{1\/2})X^{(m+1,n)}+(n+{1\/2})X^{(m,n+1)} \right) \nonumber \\
     -2\left( {m+{1\/2}\/ \l} Y^{(m+1,0)}-m
Y^{(m,0)}\right){Y^{(n,0)}\/(1+2Y^{(0,0)})^2}\;\;\;\;\; m,n\in{\bf Z}
\eqn{derX}}
where we have also
used the relations for $\tilde{\xi_r}$ and $\tilde{\eta_r}$ given in eq.
\refbr{tilde}. That equation \refbr{derX} can be employed iteratively to
obtain $X^{(m,n)}$ and $Y^{(m,n)}$ for arbitrary negative integers $m,n$ is
clear from the form of its right hand side. Since, in this paper, we intend
to give explicit results only for $m,n=0, \pm1 $ the last quantities to be
computed are those for $m=n=-1$. This is accomplished by relating first
$X^{(-1,-1)}$ to known quantities via
\refbr{derX}, and then extract $Y^{(-1,-1)}$ from \refbr{YXrel}.
Once again we find that it suffices to know $X^{(m,0)}$ and $Y^{(m,0)}$
for $m=-1,0,1,2$.

Let us now present the results for $X^{(m,0)}$ and $Y^{(m,0)}$,
$m=-1,0,1,2$, derived from $\xi_r$ and $\eta_r$ by performing first the
sums and then the contour integrals in $v^{(m)T}\xi$ and $v^{(m)T}\eta$.
After some computation one finds the expressions presented in table 1.
\begin{tabbing}
$m\;\;\;\;\;\;$\=$X^{(m,0)}\;\;\;\;\;\;\;\;\;\;\;\;\;\;\;\;\;\;\;\;\;\;
\;\;\;\;\;\;\;\;\;\;\;\;\;\;\;\;\;\;\;\;\;\;\;\;\;\;\;\;\;\;\;\;\;\;\;
\;\;$\=$Y^{(m,0)}$\\
$2$\>${\l^3
\/2}{1-\l\/(1-\l^2)^{5/2}}(1+{\l\/2}-{\l^{2}\/2}-{3\/8}\l^{3})$\>${\l^3
\/2}{1+\l\/(1-\l^2)^{5/2}}(1-{\l\/2}-{\l^{2}\/2}+{3\/8}\l^{3})$\\
$1$\>${\l^2 \/ 2}{1\/ 1+\l} {1\/ \sqrt{1-\l^2}} (1+{\l \/2})$\>${\l^2
\/2}{1\/ 1-\l}{1\/ \sqrt{1-\l^2}}(1-{\l \/2}) $\\
$0$\>${1\/2}(1-\sqrt{{1-\l \/1+\l }})$\>$-{1\/2}(1-\sqrt{{1+\l \/1-\l
}})$\\ $-1$\>${1\/2}(1-{\l \/2})$\>${1\/2}(1+{\l \/2})$
\end{tabbing}
${\bf Table\;\; 1}:\;\; X^{(m,0)}=v^{(m)T}\xi$ and $Y^{(m,0)}=v^{(m)T}\eta$
for $m=-1,0,1,2$.\newline Inserting the above results into eqs. \refbr{Xs2}
and \refbr{derX} we get the remaining quantities of interest as given in
table 2.
\eq{
\begin{array}{lll}
(m,n) & X^{(m,n)} & Y^{(m,n)} \\ (1,1) & {\l^3\/2}{1\/(1-\l^2)^2}(1-{\l
\/2}+{\l^2\/2}+{\l^3\/8}) & {\l^3\/2}{1\/(1-\l^2)^2}(1+{\l
\/2}+{\l^2\/2}-{\l^3\/8}) \\ (0,1) & {\l \/2}{1-{\l \/2}\/1-\l^2} & -{\l
\/2}{1+{\l \/2}\/1-\l^2} \\ (-1,1) & {1\/2}((1+\l)\sqrt{1-\l^2}-1) &
-{1\/2}((1-\l)\sqrt{1-\l^2}-1) \\ (1,-1) & {1\/2}(1-{1\/1+\l}{1\/
\sqrt{1-\l^2}}(1-\l^2+\l^3+{5\l^4\/4})) & -{1\/2}(1-{1\/1-\l}{1\/
\sqrt{1-\l^2}}(1-\l^2-\l^3+{5\l^4\/4})) \\ (0,-1) & {1\/2}\sqrt{{1-\l\/
1+\l}}(1+{\l\/2}) & -{1\/2}\sqrt{{1+\l\/ 1-\l}}(1-{\l\/2}) \\ (-1,-1) &
{1\/2\l}(1-{\l\/2}+{\l^2\/2}+{\l^3\/8}) &
{1\/2\l}(1+{\l\/2}+{\l^2\/2}-{\l^3\/8})
\end{array}
\nonumber }
\begin{tabbing}
${\bf Table\;\; 2}:$\= $X^{(m,n)}=v^{(m)T}{1\/1+M}v^{(n)}$ and
$Y^{(m,n)}=v^{(m)T}{1\/1-M}v^{(n)}$ for $m,n=-1,0,1$, \\
\>except those listed in table 1.
\end{tabbing}
It is gratifying to verify the relation \refbr{XYL} between the X's and the
Y's obtained by letting $\l$ go to $-\l$. Strictly speaking, of course,
having established this relation independent computations of the X's and
Y's from $\xi$ and $\eta$ are not necessary and may be viewed as a mere
check on the formalism.

Finally, as explained in the beginning of this section, it is now a simple
matter to derive the expressions appearing in the exponent of the vertex
$\hat{W}_{R_1R_2}$. The ones we promised to present the answers for are
given in table 3, where we have not tabulated
$v^{(m)T}M^T{1\/1+MM^T}v^{(n)}$ since it can be seen to equal
$v^{(n)T}M{1\/1+M^{T}M}v^{(m)}$.
\eq{
\begin{array}{llll}
(m,n) & v^{(m)T}{1\/1+M^TM}v^{(n)} & v^{(m)T}{1\/1+MM^T}v^{(n)} &
v^{(m)T}M{1\/1+M^{T}M}v^{(n)} \\ (1,1) & {\l^3
\/2}{1\/(1-\l^2)^2}(1+{\l^4\/4}) & {\l^3\/2}{1\/(1-\l^2)^2} & {\l^4
\/4}{1\/(1-\l^2)^2}(1-{3\/4}\l^2) \\ (1,0) & {\l^2 \/2}{1
\/(1-\l^2)}(1-{\l^2 \/2}) & {\l^2\/2}{1\/\sqrt{1-\l^2}} & {\l^3
\/4}{1\/(1-\l^2)} \\ (0,1) & {\l^2 \/2}{1 \/(1-\l^2)}(1-{\l^2 \/2}) &
{\l^2\/2}{1\/\sqrt{1-\l^2}} & {\l^3 \/4}{1\/\sqrt{1-\l^2}} \\ (-1,1) & {\l
\/2}{1 \/\sqrt{1-\l^2}}(1-{3\/2} \l^2) &
{\l\/2}{1\/\sqrt{1-\l^2}}(1-{3\/2}\l^2) & {1\/2}(1-{1
\/\sqrt{1-\l^2}}(1-{\l^2 \/2})^2) \\ (0,0) & {\l\/2} & {\l\/2} &
{1\/2}(1-\sqrt{1-\l^2}) \\ (1,-1) & {\l \/2}{1 \/\sqrt{1-\l^2}}(1-{3\/2}
\l^2) & {\l\/2}{1\/\sqrt{1-\l^2}}(1-{3\/2}\l^2) & {1\/2}({1
\/\sqrt{1-\l^2}}-1) \\ (-1,0) & {1\/2}\sqrt{1-\l^2} & {1\/2}(1-{\l^2\/2}) &
{\l \/4}\sqrt{1-\l^2} \\ (0,-1) & {1\/2}\sqrt{1-\l^2} & {1\/2}(1-{\l^2\/2})
& {\l \/4} \\ (-1,-1) & {1\/2\l} & {1\/2\l}(1+{\l^4\/4}) & {1\/4}(1-{3\l^4
\/4})
\end{array}
\nonumber}
${\bf Table\;\; 3}:$ Quantities appearing in the exponent of
$\hat{W}_{R_1R_2}$.\newline

These expressions are interesting also for the reason that they seem, at
least in some cases, to relate to generating functions for Catalan numbers
which appear frequently in combinatorics and statistical mechanics
problems. This observation was made in \cite{NS1} where the equation
\eq{v^T{1\/1-H^2}v={1\/\l}(1-\sqrt{1-\l^2})}
(recall that $H$ is the antisymmetric part of the matrix $M$) is discussed
and used to derive a new highly non-linear relation between binary Catalan
numbers. The connection to the Catalan numbers $c_k$ stems from the fact
that their generating function,
\eq{ S(x)={1-\sqrt{1-4x} \/2x}=\sum_{k=0}^{\infty}c_k x^k,}
obeys the algebraic equation
\eq{xS^2-S+1=0}
{}From this ones concludes that $c_0=1$ and
\eq{c_k={1\/k+1}\left(\begin{array}{c} 2k \\ k \end{array}\right)\;\;\; ,
\;\;\;k\geq1 }
It is not clear to us what the combinatorial reason behind the occurence of
Catalan numbers in this context is, and it would probably be worthwhile to
try to gain some understanding of Ramond vertices from this point of view.
For the benefit of the reader we quote here the non-linear relation found
in \cite{NS1}:
\eq{c_{k}=\sum_{m=0}^{\left[{k\/2}\right]}2^{2(m-k)}\sum_{n_{\m}}
\br{\prod_{\m=1}^{2m}{n_{\m-1}-n_{\m}\/n_{\m-1}+n_{\m}+1}}
\br{\prod_{\m=0}^{2m}(n_{\m}+1)c_{n_{\m}}}^{2}
\eqn{theCatalanformula}}
where $ n_{\m}\in\bf{Z}_{+}$ are restricted by $
\sum_{\m=0}^{2m}n_{\m}=k-m$ and where $[x]$ denotes the integer part of
$x$.

We conclude this section by briefly explaining how to arrive at an answer
for $det(1+M^TM)$. First we relate it to $det(1-M^2)$ which was computed in
\cite{CGOS}. The relation between these two determinants can be obtained by
applying the same trick as was used in \cite{CGOS} to derive a relation
between $det(1-H^2)$ and $det(1-M^2)$. First note that eq. \refbr{MT}
implies that $1+M^TM=1-M^2+2vv^TM$. If we divide this equation by $1-M^2$
and then take the determinant of it we get
\eq{det(1+M^TM)=\left( 1+2v^T{M\/1-M^2}v \right)det(1-M^2)}
Using the results in \cite{CGOS} we can thus summarize the answers for
these determinants as follows
\eqs{det(1+M^TM)&=& (1-\l^2)^{-{1\/4}}=det(1+MM^T) \eqn{detMTM} \\
    det(1-M^2)&=& (1-\l^2)^{{1\/4}} \eqn{detMM} \\ det(1-H^2)&=&
{1\/2}\left( (1-\l^2)^{-{1\/4}}+(1-\l^2)^{{1\/4}} \right) \eqn{detHH}} We
will have reason to return to these equations in the next section.

\section{The closed form of the complex four-Ramond reggeon vertex}

The purpose of this section is to propose a closed form of the four-Ramond
reggeon vertex discussed from the point of view of sewing in the previous
sections of this paper. As will be obvious below, this closed form
reproduces all our explicit results obtained in section 3 in a very direct
manner.  It has a quite natural form dictated by the structure of branch
cuts generated by the twisted fields present, and reads (for $d=1$)
\eqs{&&\hat{W}_{R_1R_2}(\l)=\bra{0}{}:\hat{\barS}(\infty) \hat{S}(-{1\/\l}):
\;:\hat{\bar{S}}(-\l)\hat{S}(0): \ket{0}{} \times \lb
 &&  : \exp\oint_{C_z}dz \oint_{C_w}dw\hat{\bar{\psi}}_R^{V^{-1}}(z)
{\bra{0}{}\hat{\psi}(z)
\hat{\bar{\psi}}(w):\hat{\bar{S}}(\infty)\hat{S}(-{1\/\l}):\;
:\hat{\bar{S}}(-\l)\hat{S}(0):
\ket{0}{}\/\bra{0}{}:\hat{\barS}(\infty) \hat{S}(-{1\/\l}):
\;:\hat{\bar{S}}(-\l)\hat{S}(0): \ket{0}{}} \hat{\psi}_R^{V^{-1}}(w):
\eqn{Wclosed}}
where radial ordering of the operators is understood in all correlation
functions. The $\hat{\psi}^{V^{-1}}_R$'s appearing in the exponent are sums
of the two external twisted fields transformed with their respective
projective transformation. The contours $C_{z}$ and $C_{w}$ enclose all
branch cuts appearing in $\hat{\barpsi}^{V^{-1}}_{R}(z)$ and
$\hatpsi^{V^{-1}}_{R}(w)$, respectively.  The normal ordering outside the
exponential refers only to these fields, while the normal ordering (also
denoted by double dots) inside the correlation functions refers to the
chiral boson in terms of which all spin fields $\hat{S}(z)$,
$\hat{\bar{S}}(z)$, and fermion fields $\hat{\bar{\psi}}(z)$,
$\hat{\psi}(z)$ are bosonized. We have chosen to normal order separately
each pair of spin fields related to the same vertex, but it should be
clear from the correlators in the exponent that the denominator can be
eliminated simply by the normal ordering of all four spin fields in the
numerator. On the other hand,
one also easily sees that by keeping the
denominator one can in fact remove all normal ordering dots in the
correlation functions in the exponent of \refbr{Wclosed}.

The exponential factor of this vertex gives, as is readily verified (see
below), exactly the results quoted in table 3 for all terms containing only
twisted zero and/or level one modes, while the prefactor correlation
function reproduces the determinant in equation \refbr{detMTM}.  There can
be hardly any doubt that this will extend to all higher modes, and we have
therefore managed to show, without providing a completely rigorous proof,
that the vertex given above (in this section) is the correct answer for the
sewed vertex with four external $R$ legs.  We will come back to the
issue of a rigorous proof in a future publication.

The form of the vertex given here, i.e. in terms of correlation functions
of spin fields etc, was suggested by LeClair in \cite{LeCl2} for the case
of one pair of twisted ($R$) external fields (one branch cut) and one
untwisted ($NS$) one, while the generalization to an arbitrary number of
$NS$ and $R$ external legs was suggested in \cite{DHMR} based on path
integral arguments, which, however, cannot in any obvious way be
transferred into the operator methods used in this paper.  In ref.
\cite{DHMR} the authors also performed the infinite dimensional
integrals giving
the NSR vertex corresponding to our eq. \refbr{matrixW} (see also the next
section where our version of their result is given and the differences
explained) but the check of the closed form was restricted to the one term
in the exponent containing only zero modes.  This term is a special
combination of $X^{(0,0)}$ and $Y^{(0,0)}$ (see section 3) and equals the
entry for $(m,n)=(0,0)$ in table 5.  The work presented in section 3 then
essentially extends these old results to all terms in the exponent by
giving an algorithm for how they may be computed, and with explicit
expressions presented for all terms containing only zero and/or level one
modes.

To get a feeling for the closed form of the reggeon vertex given in
\refbr{Wclosed}, let us evaluate the correlation function in the exponent:
\eqs{&&{\bra{0}{}\hat{\psi}(z) \hat{\bar{\psi}}(w):
\hat{\bar{S}}(\infty)\hat{S}(-{1\/\l}):\; :\hat{\bar{S}}(-\l)\hat{S}(0):
\ket{0}{}\/\bra{0}{}:\hat{\barS}(\infty) \hat{S}(-{1\/\l}):\;:
\hat{\bar{S}}(-\l)\hat{S}(0): \ket{0}{}}=\lb
&&=\bra{0}{}\hat{\psi}(z)
\hat{\bar{\psi}}(w):\hat{\bar{S}}(\infty)\hat{S}(-{1\/\l})
\hat{\bar{S}}(-\l)\hat{S}(0):\ket{0}{}=
{\sqrt{{z+{1\/\l}\/w+{1\/\l}}{z\/{z\/\l}+1}{{w\/\l}+1\/w}}\/z-w}
\eqn{feeling}}
The integrals in the exponent of \refbr{Wclosed} are then trivial to
perform because the Ramond fields multiplying the square root contain half
integer powers of the very same factors that appear under the square root,
and hence the contours $C_{z}$ and $C_{w}$ encircle in fact only poles and
no branch cuts. If external $NS$ legs are added to the surface, one
therefore has to keep their emission points away from the branch points.
In the last section, we will return to the question of how the above form
of the vertex can be generalized to an arbitrary number of external $R$ as
well as $NS$ legs.

If one rewrites equation \refbr{feeling} and the determinant prefactor
using the projective transformations $V_{i}^{-1}$ then one finds the
following alternative closed form of the four-Ramond vertex:
\eq{\hat{W}_{R_1R_2}(V_1,V_2)=
{\left({V^{-1}_2(z_1^{(1)})\/V_2^{-1}(z_2^{(1)})}\right)} ^{{1\/4}}
 :\;exp{\left(\oint_{C_z}dz\;\oint_{C_w}dw\;
\hat{\psi}^{V^{-1}}_R(z)G(V_{1},V_{2};z,w)
\hat{\bar{\psi}}^{V^{-1}}_R(w)\right)}\;:
\eqn{closedformV}}
where the propagator $G(V_{1},V_{2};z,w)$ is given by
\eq{G(V_{1},V_{2};z,w)={\sqrt{{V^{-1}_1(w)\;V^{-1}_2(w)\/V^{-1}_1(z)
\;V^{-1}_2(z)}}\/z-w}
\eqn{defofG}}
This propagator may be compared to $M(z_1,z_2;z,w)$ appearing in the
exponent of dual twisted vertex $\hat{W}_{R}(V)$ (see eqs.
\refbr{dualcomplexReqn},\refbr{M},\refbr{Mbar}) which is actually not a
propagator since it is regular at $z=w$.  Nevertheless, these two objects
are so similar that it must be considered rather strange that a simple way
of relating them does not seem to exist.  Note that in the prefactor the
role of the two projective transformations can be interchanged without
altering the result.  As we will see in the next section using other
choices of projective transformations in the sewing calculation will give
rise to closed forms of the four-vertex which if written in terms of spin
fields differ from the one presented in the beginning of this section. Not
surprisingly, however, the form given in \refbr{closedformV} is universal
to all cases (i.e. choices of projective transformations) studied here.

\section{General forms of the four-vertex for real and complex fields}

In this section we will discuss, for both real and complex fields, the
dependence of the sewed four-vertex on the choice of projective
transformations. When we described in the beginning of section 3 how to
derive the four-reggeon vertex for complex fermions by sewing we picked two
specific projective transformations $V_1$ and $V_2$. These may, however, be
chosen arbitrarily as long as they together contain only one moduli
parameter.  Here we will restrict our discussion to three other related
choices
corresponding to replacing either $V_1$, $V_2$ or both by $V_{1}\G$ and
$V_{2}\G$, respectively, where $\G(z)=1/z$. We do not attempt, however, to
perform the sewing with more general projective transformations than $V_1$
and $V_2$ since it seems that many of the tricks used in section 3 are then
no longer applicable. This is due to the fact that they depended on the
close resemblance between the matrices $M^{(\pm\pm)}(V_{i})$ that results
from the choices of $V_{1}$ and $V_{2}$ made here.

For these new choices the corresponding results from sewing are all of the
same form as the one found in the previous section, and the closed forms
differ only by moving around the bar on the bosonized spin fields in the
correlation functions appearing in \refbr{Wclosed}. We summarize this as
follows:
\eqs{&&\hat{W}_{R_{1},R_{2}}(\l)=\det{(1+M^{(++)}(V_{1})M^{(--)}(V_{2}))}:
\exp{(\sum_{i,j=1,2}\sum_{m,n}\hatpsi^{R_{i}}_{m}A_{mn}^{ij}(\l)\hat{\barpsi}
^{R_{j}}_{n})}:= \eqn {fourramondallcases} \\
&&=\bra{0}{}:\barS(z_{2}^{(1)})S(z_{1}^{(1)})::\barS(z_{2}^{(2)})
S(z_{2}^{(1)}):\ket{0}{} \eqn{closedformallcases} \\
&&:\exp{(\oint_{C_{z}}dz\oint_{C_{w}}dw\hat{\barpsi}^{V^{-1}}_{R}(z)
\bra{0}{}\psi(z)\barpsi(w):\barS(z_{2}^{(1)})S(z_{1}^{(1)})\barS(z_{2}^{(2)})
S(z_{1}^{(2)}):\ket{0}{}\hatpsi^{V^{-1}}_{R}(w))}:\nonumber}
where we have defined the infinite dimensional matrices $A^{ij}$ through
\eq{
\begin{array}{l}
A^{11}_{mn}=\sum_{r,s}U^{(+)}_{rm}(V_{1})[M^{(--)}(V_{2}){1\/1+
M^{(++)}(V_{1})M^{(--)}(V_{2})}]_{rs}\barU^{(+)}_{sn}(V_{1})\\
A^{22}_{mn}=\sum_{r,s}U^{(-)}_{rm}(V_{2})[M^{(++)}(V_{1}){1\/1+
M^{(--)}(V_{2})M^{(++)}(V_{1})}]_{rs}\barU^{(-)}_{sn}(V_{2})\\
A^{12}_{mn}=-\sum_{r,s}U^{(+)}_{rm}(V_{1})[{1\/1+
M^{(--)}(V_{2})M^{(++)}(V_{1})}]_{rs}\barU^{(-)}_{sn}(V_{2})\\
A^{21}_{mn}=\sum_{r,s}U^{(-)}_{rm}(V_{2})[{1\/1+
M^{(++)}(V_{1})M^{(--)}(V_{2})}]_{rs}\barU^{(+)}_{sn}(V_{1})
\end{array}
\eqn{theexpquantities}}
and where the four cases discussed above are given by
\eq{
\begin{array}{lllll}
{}&z_{1}^{(1)},z_{2}^{(1)}\;\;\;\;\;\;\; &
V^{-1}_{1}(z)=a{z-z_{1}^{(1)}\/z-z_{2}^{(1)}}\;\;\;\;\;\;\; &
U^{(+)}_{rm}(V_{1})=\barU^{(+)}(V_{1})\;\;\;\;\;\;\; & M^{(++)}(V_{1}) \\
{}&z_{1}^{(2)},z_{2}^{(2)} & V^{-1}_{2}(z)=a{z-z_{1}^{(2)}\/z-z_{2}^{(2)}}
& U^{(-)}_{rm}(V_{2})=\barU^{(-)}(V_{2}) & M^{(--)}(V_{2}) \\ {}&{} & {} &
{} & {} \\ I&-{1\/\l},\infty & z+{1\/\l} & \sqrt{2}v^{(m)}_{r}(\l) & M(\l)
\\ {}&0,-\l & {z\/{z\/\l}+1} & -\sqrt{2}v^{(-m)}_{r}(\l) & M^{T}(\l) \\ {}&{}
& {} & {} & {} \\ II&\infty,-{1\/\l} & {1\/z+{1\/\l}} &
\sqrt{2}iv^{(-m)}_{r}(\l) & -M^{T}(\l) \\ {}&-\l,0 & {{z\/\l}+1\/z} &
-\sqrt{2}iv^{(m)}_{r}(\l) & -M(\l) \\ {}&{} & {} & {} & {} \\
III&-{1\/\l},\infty & z+{1\/\l} & \sqrt{2}v^{(m)}_{r}(\l) & M(\l) \\
{}&-\l,0 & {{z\/\l}+1\/z} & -\sqrt{2}iv^{(m)}_{r}(\l) & -M(\l) \\ {}&{} & {}
& {} & {} \\ IV&\infty,-{1\/\l} & {1\/z+{1\/\l}} & \sqrt{2}iv^{(-m)}_{r}(\l)
& -M^{T}(\l) \\ {}&0,-\l & {z\/{z\/\l}+1} & -\sqrt{2}v^{(-m)}_{r}(\l) &
M^{T}(\l)
\end{array}
\eqn{thefourcases}}

For cases $III$ and $IV$ the results of the calculation of the matrix
elements at level
$0$ and $1$ in \refbr{fourramondallcases} using the sewing
methods described in
section three are collected in table 4.

The four complex situations described above can still be summarized in one
general closed formula for the vertex, namely
\refbr{closedformV}, the only difference being the choice of projective
transformations made in
these cases.

\eq{\begin{array}{lll}(m,n) & v^{(m)T}{M\/1-M^{2}}v^{(n)} &
v^{(m)T}{1\/1-M^{2}}v^{(n)} \\
(1,1) &{\l^4\/4}{1\/(1-\l^2)^2}(1-{\l^2\/4})
&{\l^3\/2}{1\/(1-\l^2)^2}(1+{\l^2\/2}) \\ (1,0)
&{\l^3\/4}{1\/(1-\l^2)^{3\/2}} &{\l^2\/2}{1\/(1-\l^2)^{3\/2}}(1-{\l^2\/2})
\\ (0,1) &{\l^3\/4}{1\/(1-\l^2)} &{\l^2\/2}{1\/(1-\l^2)} \\ (-1,1)
&{1\/2}(1-\sqrt{1-\l^2})&{\l\/2}\sqrt{1-\l^2} \\ (0,0)
&{1\/4}(\sqrt{{1+\l\/1-\l}}+\sqrt{{1-\l\/1+\l}}-2)
&{1\/4}(\sqrt{{1+\l\/1-\l}}-\sqrt{{1-\l\/1+\l}}) \\ (1,-1)
&{1\/2}({1\/(1-\l^2)^{3\/2}}(1-{\l^2\/2})^2-1)
&{\l\/2}{1\/(1-\l^2)^{3\/2}}(1-2\l^2+{5\l^4\/4}) \\ (-1,0) &{\l\/4} &{1\/2}
\\ (0,-1) &{\l\/4}{1\/\sqrt{1-\l^2}} &{1\/2}{1\/\sqrt{1-\l^2}}(1-{\l^2\/2})
\\ (-1,-1) &{1\/4}-{\l^2\/16} &{1\/2\l}+{\l\/4} \end{array}}

${\bf Table\;4}$: Level $0$ and $1$ matrix elements in the complex cases
$III$ and $IV$.

Having derived these complex four-vertices it is now easy to write down the
answer also for the case of real fermions taking into account the fact
that, as is usually done, the zero mode is not included in the normal
ordering (for $D$ real fields):
\eqs{&&\hatW_{R_{1},R_{2}}=\lb &=&\det{(1-H^{2})}^{D\/2}\times
\eqn{realcasesewingresult}\\
&&{}^{\times}\!\!\!\!_{\times}
e^{-\half\hatpsi^{\m T}_{R_{1}}U^{(+)}(V_{1}){H\/1-H^{2}}\;U^{(+)}(V_{1})
\hatpsi^{\m}_{R_{1}}+
\half\hatpsi_{R_{2}}^{\m T}U^{(-)}(V_{2}){H\/1-H^{2}}U^{(-)}(V_{2})
\hatpsi^{\m}_{R_{2}}+\hatpsi^{\m T}_{R_2}U^{(-)}(V_{2}){1\/1-H^{2}}
U^{(+)}(V_{1})
\hatpsi^{\m}_{R_{1}}}{}^{\times}\!\!\!\!_{\times}\;=\lb
&=&\bra{0}{}:\br{\hat{\barS}\hat{S}}(\infty,-{1\/\l}):
\;:\br{\hat{\bar{S}}\hat{S}}(-\l,0):
\ket{0}{}^{{D\/2}} \times \lb &&{}^{\times}\!\!\!\!_{\times}
\exp(\half\oint_{C_z}dz \oint_{C_w}\hatpsi_R^{\m V^{-1}}(z)
{\scriptstyle {\bra{0}{}\br{\hat{\psi}\hat{\bar{\psi}}}(z,w):
\br{\hat{\bar{S}}\hat{S}}(\infty,-{1\/\l}):\;
:\br{\hat{\bar{S}}\hat{S}}(-\l,0):
\ket{0}{}\/\bra{0}{}:\br{\hat{\barS} \hat{S}}(\infty,-{1\/\l}):
\;:\br{\hat{\bar{S}}\hat{S}}(-\l,0):
\ket{0}{}}} \hat{\psi}_R^{\m V^{-1}}(w))
{}^{\times}\!\!\!\!_{\times}\;=\eqn{realcaseclosedform}\\ &=&
( {\br{{V^{-1}_2(z_1^{(1)})\/V_2^{-1}(z_2^{(1)})}}^{1\/4}+
\br{{V^{-1}_2(z_2^{(1)})\/V_2^{-1}(z_1^{(1)})}}^{{1\/4}}\/2})^{{D\/2}}
\times \lb && {}^{\times}\!\!\!\!_{\times}
\exp\left(\half\oint_{C_{z}}dz\oint_{C_{w}}dw
\hatpsi_{R}^{\m V^{-1}}(z)K(V_{1},V_{2};z,w)
\hatpsi_{R}^{\m V^{-1}}(w)\right)
{}^{\times}\!\!\!\!_{\times} \eqn{realcaseusingprojtrans}} The results from
the sewing are given in equation \refbr{realcasesewingresult}, where
$H=\half(M-M^{T})$ and where $U^{(\pm)}$ is defined in
\refbr{defofumatrix}.  The explicit result for the determinant is given in
\refbr{detHH}, while the explicit results for level $0$ and $1$ calculated
using the method explained in section three are collected in table 5.
The closed forms, which are given in equations \refbr{realcaseclosedform}
and \refbr{realcaseusingprojtrans}, are obtained by substituting in
\refbr{Wclosed} each pair of spin fields by their symmetric combinations,
i.e. instead of $\hat{\barS}(z_{2}^{(i)}))\hat{S}(z_{1}^{(i)})$ we use, as
already noticed in \cite{DHMR},
\eq{\br{ \hat{\barS}\hat{S}}(z_{2}^{(i)},z_{1}^{(i)})=
\half(\hat{\barS}(z_{2}^{(i)})\hat{S}(z_{1}^{(i)})+
\hat{S}(z_{2}^{(i)})\hat{\barS}(z_{1}^{(i)}))\;\;\;,}
The propagator in this case becomes
\eq{K(V_{1},V_{2};z,w)=\half{1\/z-w}{
\br{{V^{-1}_2(z_1^{(1)})\/V_2^{-1}(z_2^{(1)})}}^{1\/4}
\sqrt{{V_{1}^{-1}(z)\/V_{1}^{-1}(w)}{V_{2}^{-1}(z)\/V_{2}^{-1}(w)}}+
\br{{V^{-1}_2(z_2^{(1)})\/V_2^{-1}(z_1^{(1)})}}^{1\/4}
\sqrt{{V_{1}^{-1}(w)\/V_{1}^{-1}(z)}
{V_{2}^{-1}(z)\/V_{2}^{-1}(w)}}+(z\leftrightarrow w)
\/\br{{V^{-1}_2(z_1^{(1)})\/V_2^{-1}(z_2^{(1)})}}^{1\/4}+
\br{{V^{-1}_2(z_2^{(1)})\/V_2^{-1}(z_1^{(1)})}}^{{1\/4}}}
\eqn{propagatorK}}
Note that unlike the complex case the denominator in the exponent of
\refbr{realcaseclosedform} does not cancel the correlations between the
spin fields in the numerator. Another difference to the complex case is that
the propagators in \refbr{realcasesewingresult}, \refbr{realcaseclosedform}
and \refbr{realcaseusingprojtrans}
are the same in all four cases in
\refbr{thefourcases}, since under the interchanges
$z_{1}^{(i)}\leftrightarrow z_{2}^{(i)}$ the matrices
$H^{(\pm\pm)}(z_{1}^{(i)},z_{2}^{(i)})$ and the symmetric combinations
$\br{ \hat{S}\hat{S}}(z_{1}^{(i)},z_{2}^{(i)})$ are invariant.

\eq{\begin{array}{lll} (m,n) & v^{(m)T}{1\/1-H^{2}}Hv^{(n)} &
v^{(m)T}{1\/1-H^{2}}v^{(n)} \\

(1,1) & 0 &
{\l^3\/4}{1\/(1-\l^2)^2}\left[1+{5\l^2\/4}+\sqrt{1-\l^2}(1-{\l^4\/4})\right]
\\ (0,1) & -{\l\/4}{1\/1-\l^{2}}(1-\sqrt{1-\l^{2}})^{2} &
{\l^2\/4}{1\/(1-\l^2)}\left[1+\sqrt{1-\l^2}\right] \\ (-1,1) &
{1\/2}{1\/(1-\l^2)}\left[1-\l^2-{\l^4\/8}-\sqrt{1-\l^2}(1-{\l^2\/2})\right]
& {\l\/4}{1\/(1-\l^2)}\left[1-{3\l^2\/4}+\sqrt{1-\l^2}(1-{9\l^2\/4})\right]
\\ (0,0) & 0 & {1\/\l}(1-\sqrt{1-\l^2}) \eqn{hh00}\\ (-1,0) &
-{1\/4\l}(1-\sqrt{1-\l^{2}})^{2} & {1\/4}(1+\sqrt{1-\l^{2}}) \\ (-1,-1) & 0
&
{1\/4\l}\left[1+{5\l^2\/4}+\sqrt{1-\l^2}(1-{\l^2\/4})\right]
\end{array}\nonumber}

${\bf Table\;5}$: Level $0$ and $1$ matrix elements in the real case.

These results are also obtainable by means of overlap techniques as
demonstrated in \cite{FW2}. However, when comparing the structure of the
vertex given in \refbr{Wclosed} to the corresponding expressions for real
RNS fermions obtained in \cite{FW2,DHMR} one should be aware of the fact
that our vertex is not a state in the tensor product of four Hilbert spaces
as is the case for the vertices in these references, but rather an operator
in two Hilbert spaces. This feature makes it possible to view the whole
vertex as a stringy extension of the tensor product of two Dirac matrices.
Using coherent states one can, however, rewrite our vertex in the form of a
state in the tensor product of four Hilbert spaces and thereby producing
two extra terms in the exponent.  These extra terms appear explicitly in
e.g. \cite{DHMR}.

\section{Conclusions}

In this paper we have sewn together two dual Ramond reggeon vertices to
obtain an expression for the reggeon vertex with four external R states.
Since we are using dual vertices \cite{NTWH,NT1,ENS1} the so obtained
four-vertex is an operator in the tensor product of two Hilbert spaces. One
unusual feature of the dual vertex used here is that it is normal ordered
by means of two normal ordering fields (both of $NS$ type) and is as a
consequence universal for all choices of explicit \footnote{By explicit in
this context we mean normal ordering prescriptions which do not involve any
extra fields for their definition.} normal ordering and independent of
whether we use real or complex fields.  The actual sewing, i.e. performing
of the infinite dimensional integral that results from insertion of
coherent states, does depend on whether we use complex or real fields, and
even on the specific form of the projective transformations used in
defining the emission points. Results are presented for both complex and
real fermions, and for four choices of projective transformations which
makes the computation of the four-vertex tractable. We have not tried to
carry out these calculations for more general transformations, but since
the tricks of section 3 rely heavily on the choices made here (and which is
basically the same as the ones used in all previous works on the subject),
one might anticipate problems for any other choice of transformations.

The sewing produces twisted reggeon four-vertices which are expressed in
terms of certain quantities that are built out of infinite dimensional
matrices and vectors. These quantities depend on which fermionic oscillator
modes they multiply in the exponent of the vertex and become increasingly
complicated as one considers terms with higher and higher modes. So far, to
our knowledge, only the quantity related to two zero modes has been
derived. Here we show that these quantities can, in fact, be computed at
any mode level and we present explicit formulae for all those multiplying
level zero and/or level one oscillators. These results are then compared to
a closed form of the four-vertex the structure of which was first proposed
in the case of three-vertices by LeClair \cite{LeCl2}.  The closed form of
the four-vertex, discussed also in \cite{DHMR} but there it was verified to
be correct only for the zero mode term, is easily seen to reproduce all our
explicit sewing results and there can be no doubt that this situation will
prevail for all modes.  One of the nice features about this kind vertices
is the way the branch cuts produced by the external twisted fields are
implemented.  This is done by expressing the vertex as the exponential of a
double integral over two fermi fields (each one being a sum over all
external (twisted or untwisted) fields present) connected by a propagator.
This propagator is constructed by means of a correlation functions of spin
fields and ordinary fermions distinct from the external ones. The spin
fields in these correlators create the cuts necessary for the propagator to
represent the cut surface correctly. The exact form of the vertex depends
on the projective transformations chosen, at least in the complex case.
However, a universal expression is obtained if the vertex is rewritten in
terms of the projective transformations themselves as in eq.
\refbr{closedformV}. It reads
\eqs{\hat{W}_{R_1R_2}(V_1,V_2)=&&
{\left({V^{-1}_2(z_1^{(1)})\/V_2^{-1}(z_2^{(1)})}
{V^{-1}_1(z_1^{(2)})\/V_1^{-1}(z_2^{(2)})}\right)} ^{{d\/8}}
 \lb
&&\times :\;exp{\left(\oint_{C_z}dz\;\oint_{C_w}dw\;
\hat{\psi}^{a\;V^{-1}}_R(z){\sqrt{{V^{-1}_1(w)
\;V^{-1}_2(w)\/V^{-1}_1(z)\;V^{-1}_2(z)}}\/z-w}
\hat{\bar{\psi}}^{a\;V^{-1}}_R(w)\right)}\;:
\eqn{closedformVV}}
where we have utilized the observation that
${V^{-1}_2(z_1^{(1)})\/V_2^{-1}(z_2^{(1)})}=
{V^{-1}_1(z_1^{(2)})\/V_1^{-1}(z_2^{(2)})}$
to rewrite the prefactor in a more symmetric fashion.  This suggests of
course a very direct extension to any number of twisted or untwisted
external legs. First, the fermions in the exponent are then sums over all
external (twisted or untwisted) fermions and the contours enclose all
branch cuts and poles of the external fields.  Also, the prefactor as well
as the propagator in the exponent generalize in the obvious manner to an
arbitrary number of branch cuts. In terms of spin field correlation
functions the generalization is even more obvious as already observed in
\cite{DHMR}: each new branch cut introduces a new pair of
$\bar{S}(z_1)S(z_2)$ inside the correlator which then most likely gives the
correct answer for both the propagator and prefactor.  However, one should
be aware of the fact that no proofs of any results along the lines of the
ones in this paper exist for vertices with 6 or more external Ramond legs.
In fact, in order to deal with more than four external legs one must
understand how to repeat the manipulations of section 3 when more than one
moduli parameter is involved. This problem is likely to be related to the
problems that would appear were we to use transformations different from
the ones used in this paper.


\newpage
\begin{thebibliography}{999}
\bibitem{AALO}  V. Alessandrini, D. Amati, M. Le Bellac and D. Olive,
		\PREP{1}(1971) 269.
\bibitem{Mon1}	C. Montonen, \NC{19A} (l974) 69.
\bibitem{DNPSSb}P. Di Vecchia, R. Nakayama, J.L. Petersen, S. Sciuto
		and J.R. Sidenius,\\ \NP{B287}(1987) 621.
\bibitem{PST1} 	J.L. Petersen, J.R. Sidenius and A.K. Tollst\'{e}n,
               	\NP{B317}(1989) 109.
\bibitem{DFHLPS1}P. Di Vecchia, M. Frau, K. Hornfeck, A. Lerda,
		F. Pezzella and S. Sciuto,\\ \NP{B322}(1989) 317.
\bibitem{DFHLPS2}P. Di Vecchia, M. Frau, K. Hornfeck, A. Lerda,
		F. Pezzella and S. Sciuto,\\ \NP{B333}(1990) 635.
\bibitem{ENS1}  N. Engberg, B.E.W. Nilsson and P. Sundell,
		\IJMP{A7}(1992) 4559.
\bibitem{CO}	E. Corrigan and D. Olive, \NC{11A}(1972) 749.
\bibitem{SchJ1} J. Schwarz, \NP{B65}(1973) 131.
\bibitem{Cor}	E. Corrigan, \NP{B69}(1974) 325.
\bibitem{SW1}	J.H. Schwarz and C.C. Wu, \PL{B47}(1973) 453.
\bibitem{CGOS}	E. Corrigan, P. Goddard, D. Olive and R.A. Smith,
		\NP{B67}(1973) 477.
\bibitem{SW2}	J.H. Schwarz and C.C. Wu, \NP{B72}(1974) 397.
\bibitem{SW3}	J.H. Schwarz and C.C. Wu, \NP{B73}(1974) 77.
\bibitem{SchJ2} J. Schwarz, \NP{B76}(1974) 93.
\bibitem{CFa}	E. Corrigan and D.B. Fairlie, \NP{B91}(1975) 527.
\bibitem{BCO}	D. Bruce, E. Corrigan and D. Olive, \NP{B95}(1975) 427.
\bibitem{GH}	P. Goddard and R. Horsley, \NP{B111}(1976) 272.
\bibitem{CH1}	E. Corrigan and T.J. Hollowood, \NP{B303}(1988) 135.
\bibitem{CH2}	E. Corrigan and T.J. Hollowood, \NP{B304}(1988) 77.
\bibitem{DGM2}	L. Dolan, P. Goddard and P. Montague, \NP{B338}(1990) 529.
\bibitem{BPZ}	A.A. Belavin, A.M. Polyakov and A.B. Zamolodchikov,
		\NP{B241}(1984) 333.
\bibitem{Kni1}	V.G. Knizhnik, \PL{B160}(1985) 403; \\
		D. Friedan, E. Martinec and S. Shenker,
             	\NP{B271} (1986) 93.
\bibitem{NW88a}	A. Neveu and P. West, \CMP{114}(1988) 613.
\bibitem{FW2}	M.D. Freeman and P. West, \PL{B217}(1989) 259.
\bibitem{DP1}	E. D'Hoker and D.H. Phong, \RMP{60}(1988) 917.
\bibitem{Mand1}	S. Mandelstam in Proceedings of the {\bf Workshop on
		Unified String Theories}, \\Santa Barbara 1985,
		(World Scientific 1986), M.B. Green and D.J. Gross eds.
\bibitem{AGMV2} L. Alvarez-Gaum\'{e}, C. Gomez, G. Moore
		and C. Vafa,\\ \NP{B303}(1988) 455;\\
		 L. Alvarez-Gaum\'{e}, P. Nelson, C. Gomez, G. Sierra
		and C. Vafa,\\ \NP{B311}(1988/89) 333.
\bibitem{S}	S. Sciuto, \LNC{2}(1969) 411.
\bibitem{DSS}	A. Della Selva and S. Saito, \LNC{4}(1970) 689.
\bibitem{NWbos}	A. Neveu and P. West, \NP{B278}(1986) 601;
                A. Neveu and P. West \PL{179B}(1986) 235.
\bibitem{DNPS}	P. Di Vecchia, R. Nakayama, J.L. Petersen and S.Sciuto,\\
              	\NP{B282}(1987) 189.
\bibitem{NTWH}	B.E.W. Nilsson, A.K. Tollst\'{e}n, A. W\"{a}tterstam and
		P. Hermansson,  \\ \PL{B236}(1990) 417.
\bibitem{NT1}	B.E.W. Nilsson and A.K. Tollst\'{e}n, \PL{B240}(1990) 96.
\bibitem{HNTW}	P. Hermansson, B.E.W. Nilsson, A.K. Tollst\'{e}n and
		A. W\"{a}tterstam, \\ \PL{B244}(1990) 209.
\bibitem{LeCl2}	A. LeClair, \NP{B303}(1988) 189.
\bibitem{DHMR}	P. Di Vecchia, K. Hornfeck, R. Madsen and K.O. Roland,
		\PL{B235}(1990) 63.
\bibitem{NS1}	B.E.W. Nilsson and P. Sundell, G\"{o}teborg preprint ITP 92-44
(1992).


\end{thebibliography}
\end{document}